\documentclass[twocolumn,showpacs,amsmath,amssymb,pra,superscriptaddress]{revtex4-1}
\usepackage{amssymb}
\usepackage{amsmath}
\usepackage {longtable}
\usepackage{bm}

\usepackage{graphicx}
\newcommand{\ket}[1]{|#1\rangle}
\newcommand{\bra}[1]{\langle#1|}
\newcommand \be{\begin{equation}}
\newcommand \ee{\end{equation}}
\newcommand \bea{\begin{eqnarray}}
\newcommand \eea{\end{eqnarray}}
\newcommand \bse{\begin{subequations}}
\newcommand \ese{\end{subequations}}
\begin{document}

\title{Two-qubit gates using adiabatic passage of the Stark-tuned F\"orster resonances in Rydberg atoms}

\author{I.~I.~Beterov}
\email{beterov@isp.nsc.ru}
\affiliation {Rzhanov Institute of Semiconductor Physics SB RAS, 630090 Novosibirsk, Russia}
\affiliation {Novosibirsk State University, Center of Nonlinear Photonics and Quantum Technologies, 630090 Novosibirsk, Russia}
\affiliation {Novosibirsk State Technical University, 630073 Novosibirsk, Russia}

\author{M.~Saffman}
\affiliation {Department of Physics, University of Wisconsin-Madison, Madison, Wisconsin, 53706, USA}

\author{E.~A.~Yakshina}
\affiliation {Rzhanov Institute of Semiconductor Physics SB RAS, 630090 Novosibirsk, Russia}
\affiliation {Novosibirsk State University, Center of Nonlinear Photonics and Quantum Technologies, 630090 Novosibirsk, Russia}

\author{D.~B.~Tretyakov}
\affiliation {Rzhanov Institute of Semiconductor Physics SB RAS, 630090 Novosibirsk, Russia}
\affiliation {Novosibirsk State University, Interdisciplinary Quantum Center, 630090 Novosibirsk, Russia}

\author{V.~M.~Entin}
\affiliation {Rzhanov Institute of Semiconductor Physics SB RAS, 630090 Novosibirsk, Russia}
\affiliation {Novosibirsk State University, Interdisciplinary Quantum Center, 630090 Novosibirsk, Russia}

\author{S.~Bergamini}
\affiliation{The Open University, Walton Hall, MK7 6AA, Milton Keynes, UK}

\author{E.~A.~Kuznetsova}
\affiliation {Rzhanov Institute of Semiconductor Physics SB RAS, 630090 Novosibirsk, Russia}
\affiliation {Institute of Applied Physics RAS, 603950, Nizhny Novgorod, Russia }

\author{I.~I.~Ryabtsev}
\affiliation {Rzhanov Institute of Semiconductor Physics SB RAS, 630090 Novosibirsk, Russia}
\affiliation {Novosibirsk State University, Interdisciplinary Quantum Center, 630090 Novosibirsk, Russia}

\begin{abstract}
We propose schemes of controlled-Z and controlled-NOT gates with ultracold neutral atoms based on deterministic phase accumulation during double adiabatic passage of the Stark-tuned F\"{o}rster resonance of Rydberg states. The effect of deterministic phase accumulation
during double adiabatic passage in a two-level quantum system has been analyzed in detail. Adiabatic rapid passage using nonlinearly chirped pulses with rectangle intensity profile has been discussed. Nonlinear time dependence of the energy detuning from the F\"{o}rster resonance is used to achieve a high fidelity of population transfer between Rydberg states. Fidelity of  two-qubit gates has been studied with an example of the $90S+96S\to 90P+95P$ Stark-tuned F\"{o}rster resonance in Cs Rydberg atoms. 
\end{abstract}
\maketitle

\section{Introduction}
Two-qubit quantum gates are the key element of a quantum computer. In general, any quantum algorithm can be implemented using a two-qubit controlled-NOT (CNOT) gate and single-qubit rotations~\cite{Nielsen2011}. Another example is a controlled-Z (CZ) gate which can be used for universal quantum computation as well as CNOT gate.  Experimental implementation of high-fidelity two-qubit gates is a challenging task. A two-qubit gate error below $10^{-3}$ has been  demonstrated recently for single-ion qubits~\cite{Ballance2016,Gaebler2016}. Scaling trapped ion qubits to very large quantum registers remains, however, an unsolved challenge. From this point of view, ultracold neutral atoms can be more promising candidates for implementation of a scalable quantum computer~\cite{Ryabtsev2005, Saffman2010, Comparat2010, Ryabtsev2016,  Saffman2016}. Arrays of optical dipole traps can be used as quantum registers of arbitrary dimensions~\cite{Xia2015}, and the interaction of the atom qubits to perform two-qubit gates can be controlled by their temporary excitation to Rydberg states, which have large dipole moments and experience strong long-range interactions~\cite{Jaksch2000,Lukin2001, Ryabtsev2005, Saffman2010}. For example, the effect of Rydberg excitation blockade~\cite{Lukin2001} has been successfully applied in the experiment to implement a CNOT gate for ultracold neutral atoms with the fidelity above 0.73~\cite{Maller2015a}. At the same time, high-fidelity two-qubit gates with Rydberg atoms have not been demonstrated yet. 

Another approach besides Rydberg blockade to building a two-qubit gate is based on controlled phase shifts of collective states of two qubits due to interaction between Rydberg atoms~\cite{Jaksch2000,Ryabtsev2005}. The interaction strength should be adjusted to provide a certain phase shift (for example $\pi$),  during the interaction time.  This can be easily done with Stark-tuned F\"{o}rster resonances that provide fast and flexible control by manipulating the energies of Rydberg levels with an electric field~\cite{Safinya1981, Anderson1998, Mourachko1998, Westermann2006, Nipper2012, Richards2016, Kondo2016, Pelle2016, Mandoki2016, Browaeys2016}. The Rydberg levels are adjusted in such a way that one Rydberg level lies midway between two other Rydberg states of the opposite parity. Then a resonant energy transfer between Rydberg atoms initially excited to the middle state becomes possible via resonant dipole-dipole interaction. Stark-tuned F\"{o}rster resonances for two Rydberg atoms were first reported in Ref.~\cite{Ryabtsev2010}. The rf-assisted Stark-tuned F\"{o}rster resonances have been demonstrated in Refs.~\cite{Tauschinsky2008, Ditzhuijzen2009, Tretyakov2014, Yakshina2016}. 

If two Rydberg atoms are frozen in space, dipole-dipole interaction at a F\"{o}rster resonance induces the Rabi-like coherent population oscillations between collective states of these atoms~\cite{Ryabtsev2010a}. Such oscillations have been demonstrated recently for two Rb Rydberg atoms in two optical dipole traps~\cite{Ravets2014, Ravets2015}. The frequency of these collective oscillations is sensitive to variations of the interaction energy due to fluctuations of the spatial position of the atoms within the optical dipole traps. For example, a 10\% variation of the distance between the trapped atoms results in a 25\% variation of the interaction energy due to the $1/R^{3}$ dependence of the energy of dipole-dipole interaction on distance $R$ between the atoms. This can substantially increase the phase gate error. In this paper we propose to overcome this difficulty by using a double adiabatic rapid passage across Stark-tuned F\"orster resonances with a deterministic phase accumulation. This technique is closely related to Stark-chirped rapid adiabatic passage, which is based on a laser-induced Stark shift~\cite{Shore2011, Yatsenko2002}. 

\begin{center}
\begin{figure}[!t]
\center
\includegraphics[width=\columnwidth]{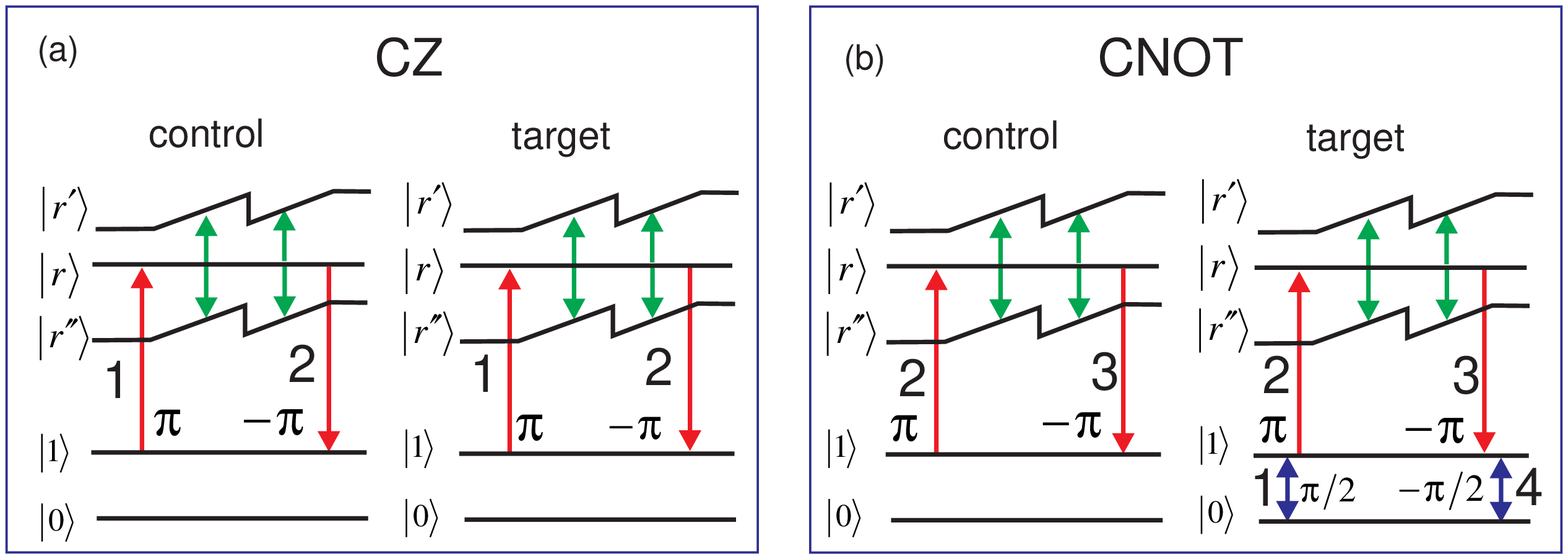}
\vspace{-.5cm}

\caption{
\label{Scheme}
(Color online) (a) Scheme of a CZ gate using double adiabatic rapid passage across Stark-tuned F\"{o}rster resonance. Two atoms are excited to  Rydberg states. An external electric field shifts the energy levels of the Rydberg atoms so that the F\"orster resonance  is passed adiabatically two times. Then the atoms are de-excited to ground state. The phase shift is deterministically accumulated if both atoms are initially prepared in state $\ket{1}$; (b) Scheme of a CNOT gate. Two additional $\pi/2$ pulses rotate the target qubit around the \textit{y} axis in the opposite directions.
}

\end{figure}
\end{center}

A scheme of CZ gate is shown in Fig.~\ref{Scheme}(a). Two optical dipole traps with one atom in each trap are located at a distance $R$ between them. The two atoms are simultaneously excited to Rydberg state $\ket{r}$ by a $\pi$ laser pulse labeled as 1. The distance between the traps must be sufficiently large to avoid the effect of Rydberg blockade~\cite{Lukin2001}. A time-dependent external electric field shifts the collective energy levels so that the F\"orster resonance $\ket{rr} \to \ket{r'r''}$ is passed adiabatically two times. This results in a deterministic phase shift of state $\ket{rr}$. After the end of adiabatic passage the atoms are de-excited to ground state by a $-\pi$ laser pulse labeled as 2. 

The phase shift due to Rydberg-Rydberg interaction is accumulated only in the case when both atoms are initially prepared in state 
$\ket{1}$ and then excited to Rydberg state $\ket{r}$. If one of the atoms (or both of them) is initially in the state $\ket{0}$, no phase shift occurs.

A scheme of CNOT gate is shown in Fig.~\ref{Scheme}(b). Two additional $\pi/2$ pulses, labeled as 1 and 4, rotate the target qubit around the \textit{y} axis in the opposite directions. If the control qubit is initially prepared in state $\ket{0}$ and is not excited to Rydberg state, the pulse sequence acting on the target qubit returns it back to the initial state. The $\pi$ phase shift due to the Rydberg-Rydberg interaction results in the inversion of the state of target qubit, if the control qubit is initially prepared in state $\ket{1}$.

The paper is organized as follows. In Sec.~III we explain the effect of deterministic phase accumulation during double adiabatic rapid passage in a two-level quantum system. In Sec.~IV we discuss the features of adiabatic rapid passage across Stark-tuned F\"{o}rster resonance for two interacting Cs Rydberg atoms with nonlinear time dependence of the detuning from the resonance. Fine structure and finite lifetimes of the Rydberg states have been taken into account in our analysis.

\section{Phase accumulation during adiabatic rapid passage}

Adiabatic rapid passage is commonly used for laser excitation of molecular levels because of the independence of transition probability on the Rabi frequency~\cite{Malinovsky2001}. A number of schemes for quantum logic using two-photon stimulated Raman adiabatic passage (STIRAP)~\cite{Bergmann1998} and Rydberg excitation has been developed~\cite{Moller2008,Rao2014}. In our previous works~\cite{Beterov2013,Beterov2014} we have found that double adiabatic rapid passage returns the system to the initial state, but with a deterministic phase shift. This shift is equal to $\pi $ for two identical laser pulses and to zero if the second laser pulse has the opposite sign of Rabi frequency. This allowed us to develop schemes of quantum gates with mesoscopic atomic ensembles, using adiabatic passage and Rydberg blockade~\cite{Beterov2013,Beterov2014}. Below we explain the effect of deterministic phase accumulation using a theory of adiabatic rapid passage~\cite{Berman2011}. The Hamiltonian for a two-level system with states $\ket{1}$ and $\ket{2}$, interacting with a chirped laser pulse (laser frequency and intensity change during the pulse), is written as

\be
\label{eq1}
\hat{\mathbf{H}}\left(t\right)=\frac{\hbar }{2} \left(\begin{array}{cc} {0} & {\Omega_0 \left(t\right)} \\ {\Omega_0 \left(t\right)} & {2\delta \left(t\right)} \end{array}\right).
\ee

\noindent Here $\Omega_0 \left(t\right)$ is time-dependent Rabi frequency and $\delta \left(t\right)$ is time-dependent detuning from the resonance. In the field interaction representation~\cite{Berman2011} the wavefunction is written as

\be
\label{eq2} 
\psi \left(t\right)=c_1\left(t\right)e^{i\omega t/2 }\ket{1} +c_{2} \left(t\right)e^{-i\omega t / 2}\ket{2}. 
\ee

\noindent Here $c_1\left(t\right)$ and $c_2\left(t\right)$ are probability amplitudes and $\omega$ is laser frequency. We define the 
time-dependent basis states to be $\ket{1\left(t\right)} =e^{i\omega t}\ket{1}$ and $\ket{2\left(t\right)} =e^{-i\omega t}\ket{2}$. In this basis the wavefunction is rewritten as follows:
\be
\label{eq3} 
\ket{\psi \left(t\right)} =c_1 \left(t\right)\ket{ 1\left(t\right)} +c_2 \left(t\right)\ket{2\left(t\right)}.
\ee

\noindent To diagonalize the Hamiltonian, we rotate the basis:
\be
\label{eq4}
\left(\begin{array}{c} {\ket{ I\left(t\right) } } \\ {\ket{ II\left(t\right) } } \end{array}\right)=\mathbf{T}\left(t\right)\left(\begin{array}{c} {\ket{ 1\left(t\right) } } \\ {\ket{ 2\left(t\right) } } \end{array}\right).    
\ee

\noindent Here $\ket{I\left(t\right)}$ and $\ket{II\left(t\right)}$ are semiclassical dressed states~\cite{Berman2011} and  $\mathbf{T}\left(t\right)$ is time-dependent unitary rotation matrix: 

\be
\label{eq5}
\mathbf{T}\left(t\right)=\left(\begin{array}{cc} {\cos \theta \left(t\right)} & {-\sin \theta \left(t\right)} \\ {\sin \theta \left(t\right)} & {\cos \theta \left(t\right)} \end{array}\right).      
\ee 

\noindent where  $\theta \left(t\right)$ is a time-dependent mixing angle. The semiclassical dressed states are the superpositions:

\be
\label{eq6}
\begin{array}{l} {\ket{ I\left(t\right) } =\cos \theta \left(t\right)\ket{ 1\left(t\right) } -\sin \theta \left(t\right)\ket{ 2\left(t\right)} } \\ {\ket{II\left(t\right) } =\sin \theta \left(t\right)\ket{ 1\left(t\right) } +\cos \theta \left(t\right)\ket{ 2\left(t\right) } } \end{array}.     
\ee

\noindent To derive the equation for the probability amplitudes of dressed states  $\tilde{\mathbf{c}}$, we substitute the definition  $\tilde{\mathbf{c}}=\mathbf{T}\mathbf{c}$ into the Schr\"{o}dinger equation for the probability amplitudes $i\hbar \mathbf{\dot{c}=\hat{H}c}$. This results in

\be
\label{eq7} 
i\hbar \dot{\tilde{\mathbf{c}}}=\mathbf{T}\hat{\mathbf{H}}\mathbf{T}^{\dagger} \tilde{\mathbf{c}}-i\hbar \mathbf{T}\dot{\mathbf{T}}^{\dagger} \tilde{\mathbf{c}}.
\ee

\noindent The matrix $\mathbf{T\hat{H}T^{\dagger}}$ is diagonal if the mixing angle $\theta\left(t\right)$ obeys the following conditions:

\be
\label{eq8}
\begin{array}{l} 
{\tan\left[2\theta \left(t\right)\right]=\Omega_{0} \left(t\right) / \delta \left(t\right)} \\ 
{\sin \left[\theta \left(t\right)\right]=\sqrt{\dfrac{1}{2} \left(1-\dfrac{\delta \left(t\right)}{\Omega \left(t\right)} \right)} } \\
 {\cos \left[\theta \left(t\right)\right]=\sqrt{\dfrac{1}{2} \left(1+\dfrac{\delta \left(t\right)}{\Omega \left(t\right)} \right)} } \end{array}.     
\ee

\noindent Here $\Omega \left(t\right)=\sqrt{\Omega _{0}^{2} \left(t\right)+\delta \left(t\right)^{2} }$. This leads to:

\be
\label{eq9}
\begin{array}{l} {\hat{\mathbf{H}}_{d} =\mathbf{T\hat{H}T^{\dagger}} =\frac{\hbar }{2} \left(\begin{array}{cc} {\Omega_{\_ }\left(t\right)} & {0} \\ {0} & {\Omega_+ \left(t\right)} \end{array}\right)} \\ {\mathbf{T\dot{T}^{\dagger}} =i\sigma_{y} \dot{\theta}} \end{array}.
\ee

\noindent Here $\Omega_{\_ }=\delta\left(t\right)-\Omega\left(t\right)$ and $\Omega_+=\delta\left(t\right)+\Omega\left(t\right)$. In the adiabatic approximation, when $\left|\dot{\Omega}_0\left(t\right)\right|/\Omega^2\left(t\right)\ll1$ and $\left|\dot{\delta}\left(t\right)\right|/\Omega^2\left(t\right)\ll1$ we can neglect the term proportional to $\dot{\theta}$. Then Eq.~(\ref{eq7}) is rewritten as $i\hbar \dot{\tilde{\mathbf{c}}}=\hat{\mathbf{H}}_{d} \tilde{\mathbf{c}}$.  Its solution is 
\be
\label{eq10}
\begin{array}{l} {\tilde{c}_{1} \left(t\right)=\tilde{c}_{1} \left(0\right)\exp \left[-\frac{i}{2}\int\limits_{0}^{t}\Omega_{\_} \left(t\right)dt \right]} \\ {\tilde{c}_{2} \left(t\right)=\tilde{c}_{2} \left(0\right)\exp \left[-\frac{i}{2}\int\limits_{0}^{t}\Omega_+ \left(t\right)dt \right]} \end{array}.    
\ee

\begin{center}
\begin{figure}[!t]
 \center
\includegraphics[width=\columnwidth]{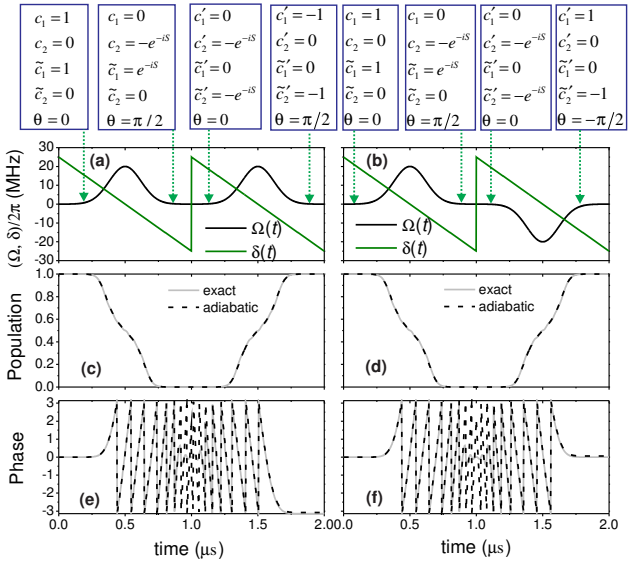}
\vspace{-.5cm}
\caption{
\label{Analyt}(Color online).
 Scheme of deterministic phase accumulation during a double adiabatic passage in a two-level quantum system. The phase shift is $\pi$ for the left-hand panel and zero for the right-hand panel. The dynamics of probability amplitudes $c_{1} \left(t\right)$,$c_{2} \left(t\right)$, $c'_{1} \left(t\right)$, $c'_{2} \left(t\right)$ of states  $\ket{1\left(t\right),2\left(t\right)} $  and of probability amplitudes $\tilde{c}_{1} \left(t\right)$, $\tilde{c}_{2} \left(t\right)$, $\tilde{c'}_{1} \left(t\right)$, $\tilde{c'}_{2} \left(t\right)$ of semiclassical dressed states $\ket{ I\left(t\right),II\left(t\right)}$ is shown schematically.  (a), (b) Time dependences of Rabi frequency $\Omega \left(t\right)$ and of detuning $\delta \left(t\right)$; (c), (d) Numerically calculated time dependences of the population of  initial state $\ket{1\left(t\right)}$ compared with the calculations in the adiabatic approximation. (e), (f) Numerically calculated time dependencies of the phase of  initial state $\ket{1\left(t\right)}$ compared with calculations in the adiabatic approximation. 
}
\end{figure}
\end{center}

\noindent Now we consider a double adiabatic sequence which starts at \textit{t}=0. The time dependence of Rabi frequency $\Omega_0\left(t\right)$ and detuning $\delta\left(t\right)$ is illustrated in Fig.~\ref{Analyt}(a). The system is initially in state $\ket{1\left(t\right)}$. For initial positive detuning $\delta \left(0\right)>0$ and $\Omega_0 \left(0\right)=0$ we find $\Omega \left(0\right)=\delta \left(0\right)$ and therefore $\theta \left(0\right)=0$. From Eq.~(\ref{eq6}) the initial dressed state is  $\ket{I\left(t\right)}$ and $\tilde{c}_{1} \left(0\right)=1$. The time-dependent probability amplitudes are 

\be
\label{eq11}
\begin{array}{l} {c_1 \left(t\right)=\tilde{c}_1 \left(t\right)\cos \theta \left(t\right)} \\ 
{c_2 \left(t\right)=-\tilde{c}_1 \left(t\right)\sin \theta \left(t\right)} \end{array}.
\ee

\noindent After the end of the first adiabatic passage at time $T$ the detuning is negative $\delta \left(T\right)<0$ and $\Omega \left(T\right)=-\delta \left(T\right)$. Therefore the mixing angle $\theta \left(T\right)=\pi /2$, and the system ends in state $\ket{2\left(t\right)}$ with $c_2 \left(T\right)=-\tilde{c}_1 \left(T\right)=-\exp\left[-\frac{i}{2}\int_0^{T}\Omega_{\_}\left(t\right)dt \right]=-e^{-iS}$. 

We denote the mixing angle and the probability amplitudes for the second adiabatic passage as $\theta'$, $c_1'\left(t\right)$, $c_2'\left(t\right)$, $\tilde{c}_1'\left(t\right)$, $\tilde{c}_2'\left(t\right)$. At the beginning of the second adiabatic passage the detuning is positive $\delta \left(T\right)>0$ and $\theta'\left(T\right)=0$. At time $t=T$ the system is in state $\ket{2\left(t\right)}$. From Eq.~(\ref{eq6}) the dressed state is now $\ket{II\left(t\right)}$. The probability amplitude $c_2\left(t\right)$ of state $\ket{2\left(t\right)}$ is constant around $t=T$ due to the absence of interaction with the laser field. Therefore, the initial probability amplitude of dressed state  $\ket{II\left(t\right)}$ is $\tilde{c}'_2 \left(T\right)=c_2\left(T\right)=-\tilde{c}_{1} \left(T\right)$. During the second adiabatic passage the time-dependent probability amplitudes are expressed similarly to Eq.~(\ref{eq11}):

\be
\label{eq12}
\begin{array}{l} {c'_1 \left(t\right)=\tilde{c}'_2 \left(t\right)\sin \theta '\left(t\right)} \\
 {c'_2 \left(t\right)=\tilde{c}'_2 \left(t\right)\cos \theta '\left(t\right)} \end{array}.
\ee

\noindent From Eq.~(\ref{eq10}) the probability amplitude of dressed state $\ket{II\left(t\right)}$ is 
$\tilde{c}'_2 \left(t\right)=\tilde{c}'_2 \left(T\right)\exp \left[-\frac{i}{2}\int _{T}^{t}\Omega_+ \left(t\right)dt \right]$. After the end of the second adiabatic passage the mixing angle is $\theta '\left(2T\right)=\pi /2$ and the system ends in state $\ket{1\left(t\right)}$ with probability amplitude 

\bea
\label{eq13} c'_{1}\left(2T\right)&=&\tilde{c}'_2 \left(2T\right)=\\
&=&-\exp \left[-\frac{i}{2}\int\limits_{T}^{2T}\Omega_{+} \left(t\right)dt \right]\exp \left[-\frac{i}{2}\int\limits_{0}^{T}\Omega_{\_} \left(t\right)dt \right]. \nonumber
\eea

\noindent For two identical laser pulses with identical time dependences of the detuning we find $c'_{1} \left(2T\right)=-1$.

This $\pi$ phase shift can be compensated if the second laser pulse has the opposite sign of Rabi frequency $\Omega _{0} \to -\Omega _{0}$ (or a $\pi$ phase shift of the laser field), as shown in Fig.~\ref{Analyt}(b). To diagonalize the Hamiltonian for the second adiabatic passage, we modify Eq.~(\ref{eq8}):

\be
\label{eq14}
\begin{array}{l} 
{\tan\left[2\theta \left(t\right)\right]=-\Omega_{0} \left(t\right) / \delta \left(t\right)} \\ 
{\sin \left[\theta \left(t\right)\right]=-\sqrt{\dfrac{1}{2} \left(1-\dfrac{\delta \left(t\right)}{\Omega \left(t\right)} \right)} } \\
 {\cos \left[\theta \left(t\right)\right]=\sqrt{\dfrac{1}{2} \left(1+\dfrac{\delta \left(t\right)}{\Omega \left(t\right)} \right)} } \end{array}.     
\ee

\noindent In this case after the end of the second adiabatic passage $\theta'\left(2T\right)=-\pi /2$ and $c'_1\left(2T\right)=1$.

To illustrate this model we have numerically calculated the time dynamics of probability amplitudes of a two-level atom interacting with two chirped laser pulses with time-dependent Rabi frequency $\Omega_{0j} \left(t\right)=\Omega_0 \exp \left[-\left(t-t_j \right)^2 / 2w^2 \right]$ and detuning  $\delta_j \left(t\right)=s_1 \left(t-t_j \right)$; where \textit{j}=1,2, as shown in Fig.~\ref{Analyt}(a). The peak Rabi frequency is $\Omega_0  / 2\pi=20$~MHz, the chirp of detuning is $s_1 /  2\pi =-50\;\mathrm{MHz} /\mu s$, and the pulse width is $w=0.12\;\mu s$. The centers of the pulses are located at the times $t_{1} =0.5\;\mu s$ and $t_{2} =1.5\; \mu s$. The conditions of Fig.~\ref{Analyt}(b) are similar, but the second pulse has the opposite sign of Rabi frequency. Figures~\ref{Analyt}(c) and \ref{Analyt}(d) show the numerically calculated time dependence of probability $P_1 =\left|c_{1} \left(t\right)\right|^{2} $ to find the system in initial state $\ket{1}$. Figures~\ref{Analyt}(e) and \ref{Analyt}(f) show the numerically calculated phase $arg\left[c_{1} \left(t\right)\right]$ of state $\ket{1}$. The exact solution of the Schr\"{o}dinger equation with the Hamiltonian from Eq.~(\ref{eq1}) is compared with the adiabatic approximation from Eqs.~(\ref{eq10})-(\ref{eq14}). Good agreement between the exact solution and adiabatic approximation is observed. Notably, the disagreement between exact and adiabatic phase dynamics in the region where $c_1 \left(t\right)\approx 0$ does not affect the subsequent behavior of the system. After  double adiabatic sequence the system returns to initial state with phase shift $\pi$ in the left-hand panel of Fig.~\ref{Analyt} and with zero phase shift in the right-hand panel of Fig.~\ref{Analyt}.

\begin{center}
\begin{figure}[!t]
 \center
\includegraphics[width=\columnwidth]{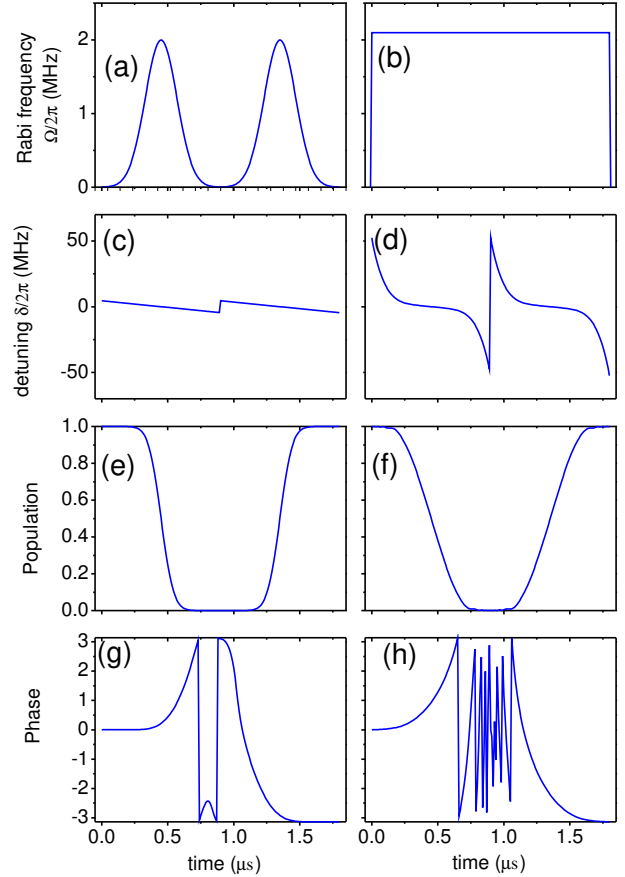}
\vspace{-.5cm}
\caption{
\label{Nonlin}(Color online).
Comparison between the schemes of double adiabatic rapid passage with linearly chirped Gaussian pulses (left-hand panel), and with rectangular shape of time-dependent Rabi frequency and nonlinear time dependence of detuning  from the resonance (right-hand panel). (a),(b) Time dependence of Rabi frequency $\Omega \left(t\right)$; (c),(d) Time dependence of detuning from the resonance $\delta \left(t\right)$; (e),(f) Time dependence of the population of state $\ket{1} $; (g),(h) Time dependence of the phase of state $\ket{1}$.
}
\end{figure}
\end{center}

\section{Adiabatic passage across Stark-tuned F\"{o}rster resonance}

The energy of dipole-dipole interaction of two Rydberg atoms is determined by the interatomic distance which cannot be changed on short timescales. Therefore we need to consider the adiabatic rapid passage with rectangular shape of the time-dependent Rabi frequency. To achieve high fidelity of the population transfer, we use a nonlinear time dependence of the detuning from the resonance:

\be
\label{eq15}
\delta _{k} \left(t\right)=s_{1} \left(t-t_{k} \right)+s_{2} \left(t-t_{k} \right)^{5}.  
\ee

\noindent Here the exact resonance occurs at the times $t_k$ with $k$=1,2. The detuning is slowly varied across the resonance and is rapidly increased before and after the resonance, which is close to the approach of high-fidelity laser driving~\cite{Bason2012}. Figure~\ref{Nonlin} illustrates the difference between the conventional scheme of adiabatic rapid passage, which uses chirped Gaussian pulses with linear time dependence of detuning (left-hand panel), and the scheme of adiabatic rapid passage with rectangular shape of the time-dependent Rabi frequency and nonlinear time dependence of detuning (right-hand panel). The parameters of the pulses for the left-hand panel of Fig.~\ref{Nonlin} are $\Omega_{0k} \left(t\right)=\Omega_0 \exp \left[-\left(t-t_k \right)^2 / 2w^2 \right]$ with $\Omega_0/2\pi=2$~MHz, $w=0.12\; \mu s$, and $\delta_k \left(t\right)=s_1 \left(t-t_{k} \right)$ with 
$s_1=-10\;\mathrm{MHz}/\mu s$. For the right-hand panel of Fig.~\ref{Nonlin} Rabi frequency is constant $\Omega_{0k}\left(t\right)/2\pi=2.1$~MHz, and the detuning is described by Eq.~(\ref{eq15}) with $s_1/2\pi=-10\;\mathrm{MHz}/\mu s$, and $s_2/2\pi=-2600\;\mathrm{MHz}/\mu s^5$. The centers of the pulses are located at times $t_1 =0.45\; \mu s$ and $t_{2} =1.35\; \mu s$. The population error for the final state of the system is found to be below $3\times 10^{-5}$ in both cases. The phase shift is equal to $\pi$ in both cases.

Stark-tuned F\"{o}rster resonance required for the implementation of the proposed scheme must meet the following criteria: (i) the lifetimes of Rydberg states must be sufficiently long to avoid the decay of coherence during the gate operation due to spontaneous and blackbody radiation (BBR) induced transitions; (ii) initial F\"{o}rster energy defect must be sufficiently large to allow for rapid turning off the interaction between atoms at the beginning and the end of the adiabatic passage; (iii) selected interaction channel must be well isolated from the other channels to avoid break-up or dephasing of the adiabatic population transfer.

\begin{center}
\begin{figure}[!t]
 \center
\includegraphics[width=\columnwidth]{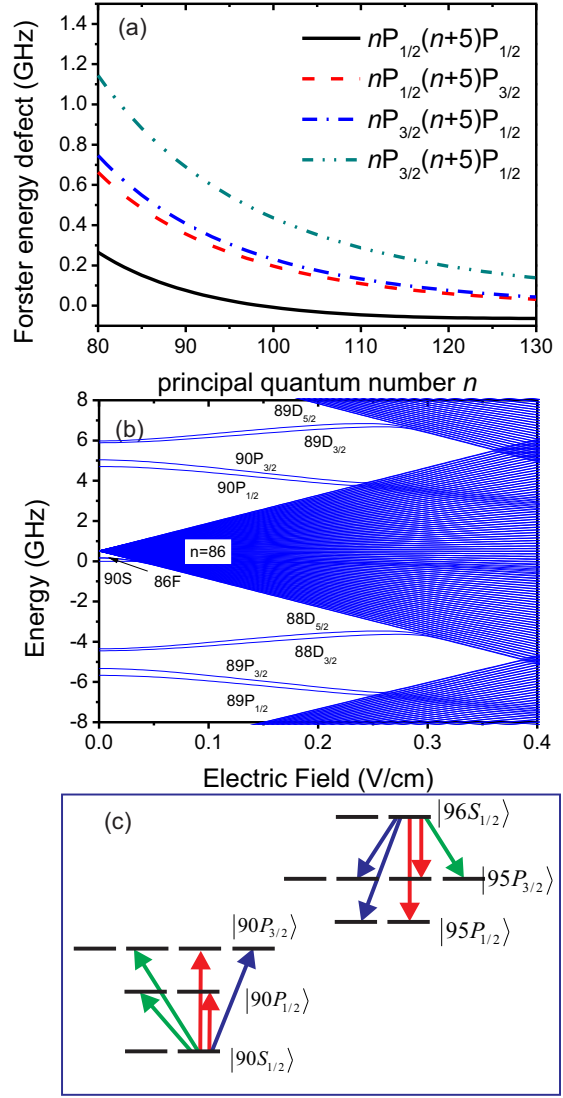}
\vspace{-.5cm}
\caption{
\label{Stark}(Color online)  (a) Energy defect of the F\"{o}rster resonance ${\left| nS,\left(n+6\right)S \right\rangle} \to {\left| nP,\left(n+5\right)P \right\rangle} $ in Cs Rydberg atoms; (b) Stark diagram for Cs Rydberg states with $\left|m_{j} \right|=1/2 $. The 90S state is selected as zero energy level; (c) Scheme of possible transitions for various channels of the ${\left| 90S,96S \right\rangle} \to {\left| 90P,95P \right\rangle} $ F\"{o}rster resonance in Cs. The \textit{nS} states with $m_{j} ={1 /2}$ are initially excited.
}
\end{figure}
\end{center}

In our previous work~\cite{Beterov2015} we have studied the structure of the F\"{o}rster resonances $\ket{nS,n'S} \to \ket{nP,\left(n'-1\right)P}$ in Rb and Cs Rydberg atoms. The  energy defect for Cs $\ket{nS,\left(n+6\right)S}\to \ket{nP,\left(n+5\right)P}$ F\"{o}rster resonance in a zero electric field is shown in Fig.~\ref{Stark}(a) for the range of principal quantum numbers $80<n<130$. We have selected the $\ket{90S_{1/2} ,96S_{1/2}} \to \ket{90P_{1/2},95P_{1/2}}$ Stark-tuned F\"{o}rster resonance for the further numerical simulations. This resonance has the energy defect  $\delta_0/2 \pi =75.6$~MHz in a zero electric field. In contrast to the resonances involving $\ket{nP_{3/2}}$ states, this resonance has no Stark splitting in the electric field.

The Stark diagram for Cs Rydberg states with $\left|m_j \right|=1/2$ is shown in Fig.~\ref{Stark}(b). The dc electric field is aligned along the \textit{z} axis. The 90$S$ state is selected as zero energy level. We have calculated the radial matrix elements using the quasiclassical approximation~\cite{Dyachkov1994} and the method of quantum defects~\cite{Lorenzen1984, Weber1987,Deiglmayr2016}. The Stark shift for \textit{nS} and \textit{nP} Rydberg states is close to quadratic and is approximated as

\be
\label{eq16}
\delta \left(E\right)=-\frac{1}{2} \alpha E^{2}.
\ee

\noindent The polarizabilities $\alpha$, listed in Table~I, have been found from the numeric approximation of the Stark energy shift for the electric field $E<50$~mV/cm. The exact F\"{o}rster resonance $\ket{90S_{1/2} ,96S_{1/2}} \to \ket{90P_{1/2},95P_{1/2}}$  occurs in the electric field $E=29.75$~mV/cm.

\begin{table}
\caption{Calculated polarizabilities of Cs Rydberg states}
\begin{tabular*}{\columnwidth}{@{\extracolsep{\fill}}|c|c|c|} \hline 
State & $\left|m_{j} \right|$ &  $\alpha \left[\frac{{\rm MHz}}{\left({{\rm V} \mathord{\left/{\vphantom{{\rm V} {\rm cm}}}\right.\kern-\nulldelimiterspace} {\rm cm}} \right)^{2} } \right]$ \\ \hline 
$\ket{90S_{1/2}} $ & 1/2 &  3505 \\ \hline 
$\ket{96S_{1/2}} $ & 1/2 &  5529 \\ \hline 
$\ket{90P_{1/2}} $ & 1/2 &  72511 \\ \hline 
$\ket{90P_{3/2}} $ & 1/2 &  103738 \\ \hline 
$\ket{90P_{3/2}} $ & 3/2 & 87196 \\ \hline 
$\ket{95P_{1/2}} $ & 1/2 &  107380 \\ \hline 
$\ket{95P_{3/2}} $ & 1/2 &  153574 \\ \hline 
$\ket{95P_{3/2}} $ & 3/2 &  129118 \\ \hline 
\end{tabular*}
\end{table}

 The operator of dipole interaction between atoms \textit{A} and \textit{B} with interatomic separation \textit{R} along the \textit{z} axis is

\be
\label{eq17} V_{dd} =\frac{e^2}{4\pi \varepsilon_{0} R^3} \left(\mathbf{a \cdot b}-3a_z b_z \right).  
\ee

\noindent Here $\mathbf{a}$ is a vectorial position of the electron in atom A and $\mathbf{b}$ is a vectorial position of the electron in atom B~\cite{Walker2008}. For the states $\ket{\gamma_a}=\ket{90S}$, $\ket{\gamma_b}=\ket{96S}$, $\ket{\gamma_{\alpha}}=\ket{90P}$ and $\ket{\gamma_{\beta}}=\ket{95P}$ the matrix elements for the dipole-dipole interaction operator are written as~\cite{Beterov2015}

\begin{eqnarray}
M_{m_a,m_b}^{m_{\alpha },m_{\beta }}&=&\frac{C_{3,k} }{R^3}Q_k \\ \nonumber
Q_k&=&-\sqrt6\sum_{q=-1}^1 C_{1q1-q}^{20}
C_{j_am_a1q}^{j_{\alpha }m_{\alpha }}C_{j_bm_b1-q}^{j_{\beta }m_{\beta }}.
\end{eqnarray}

\noindent Here $Q_k$ is the angular factor (see Table~II) which is only nonzero for $m_a+m_b=m_{\alpha}+m_{\beta}$, and the dipole-dipole interaction energy $C_{3,k}$ coefficient is expressed as

\begin{equation}
C_{3,k}(a,b,\alpha,\beta)=q^2\frac{\langle\gamma_{\alpha }||r_a||\gamma_a \rangle
\langle\gamma_{\beta }||r_b||\gamma_b\rangle}{\sqrt{(2j_{\alpha }+1)(2j_{\beta }+1)}},
\label{eq19}
\end{equation}
\noindent with $q^2=e^2/4\pi\epsilon_0$, $e$ is the electronic charge, $\epsilon_0$ is the permittivity of free space, and  $\langle\gamma_{\alpha }||r_a||\gamma_a\rangle$ is a reduced matrix element in the fine structure basis:

\bea
\label{eq20} 
&{\left\langle \gamma_{\alpha} \right|} \left|r\right|{\left| \gamma_a \right\rangle} =(-1)^{\frac{l_{\alpha}+l_a}{2}+j_a}\sqrt{\max \left(l_{\alpha} ,l_{a} \right)} \times& \\
&\times \sqrt{2j_{\alpha} +1} \sqrt{2j_a +1} \left\{\begin{array}{ccc} {l_a } & {1/2} & {j_a } \\ {j_{\alpha} } & {1} & {l_{\alpha}} \end{array}\right\}r. &\nonumber
\eea

\noindent Here \textit{r} is the radial matrix element.

In our numerical simulations we considered the F\"{o}rster resonances $\ket{90S,96S} \to \ket{nP,n'P}$ and the subsequent transitions $\ket{nP,n'P} \to \ket{mS,m'S}$, $\ket{nP,n'P} \to \ket{mD,m'D}$, $\ket{nP,n'P} \to \ket{mD,m'S}$, $\ket{nP,n'P} \to \ket{mS,m'D}$  with F\"{o}rster energy defect $\Delta_0$ less than 1~GHz and principal quantum numbers $88<n,n',m,m'<98$. We have identified 116 resonances, taking into account fine structure and Stark sublevels of the Rydberg states. Examples of such resonances are given in Table~II. 

\begin{table}
\caption{Examples of F\"{o}rster resonance channels $\ket{\gamma_a,\,\gamma_b}\to \ket{\gamma_{\alpha},\,\gamma_{\beta}}$ with F\"{o}rster energy defect  $\Delta_0$ and F\"{o}rster interaction energy $C_{3,k}$.}
\begin{tabular*}{\columnwidth}{@{\extracolsep{\fill}}|c|c|c|c|c|c|p{2cm}|} \hline 
 & $\ket{\gamma_a}$ & $\ket{\gamma_b}$ & $\ket{\gamma_{\alpha}}$ & $ \ket{\gamma_{\beta}}$ & $\Delta_0$~(MHz)& $C_{3,k}$ $ (\mathrm{MHz} \, \mu \rm m^3$)  \\ \hline 
1 & $90S_{1/2}$ & $96S_{1/2}$ & $90P_{1/2}$ & $95P_{1/2}$ & 75.610 & -154968\\ \hline 
2 & $90S_{1/2}$ & $96S_{1/2}$ & $90P_{1/2}$ & $95P_{3/2}$ & 356.525 & 162160\\ \hline 
3 & $90S_{1/2}$ & $96S_{1/2}$ & $90P_{3/2}$ & $95P_{1/2}$ & 408.152 & 149112 \\ \hline 
4 & $90S_{1/2}$ & $96S_{1/2}$ & $90P_{3/2}$ & $95P_{3/2}$ & 689.067 & -156032\\ \hline 
5 & $90S_{1/2}$ & $96S_{1/2}$ & $95P_{1/2}$ & $90P_{1/2}$ & 75.610 & -26\\ \hline
6 & $90P_{1/2}$ & $95P_{1/2}$ & $88S_{1/2}$ & $97D_{3/2}$ & -644.278 & -240\\ \hline 
7 & $90P_{1/2}$ & $95P_{1/2}$ & $88D_{3/2}$ & $95D_{3/2}$ & 288.906 & 11043\\ \hline 
\end{tabular*}
\end{table}

 We have found that only channels 1-4 from Table~II are responsible for the time dynamics of the initial collective state $\ket{90S_{1/2} , m_a=1/2 ;\, 96S_{1/2} ,m_b=1/2}$. This corresponds to eight most important F\"{o}rster interaction channels $\ket{\gamma_a\,m_a,\,\gamma_b\,m_b}\to \ket{\gamma_{\alpha}\,m_{\alpha},\,\gamma_{\beta}\,m_{\beta}}$ for different Stark sublevels, which are listed in Table~III and shown in Fig.~\ref{Stark}(c). In particular, the F\"{o}rster resonance $\ket{ 90P_{1/2},\, 95P_{1/2}} \to \ket{88D_{3/2}, \, 95D_{3/2}}$ with the detuning 288.9~MHz in zero electric field has the effect below $10^{-4}$ on the calculated probability to find the atomic system in the state $\ket{90S_{1/2} ;\, 96S_{1/2}}$ and its phase after the end of the adiabatic passage. Therefore in our simulations of the CNOT gate this channel and other transitions to $D$ and $S$ states were not taken into account.

\begin{center}
\begin{figure}[!t]
 \center
\includegraphics[width=\columnwidth]{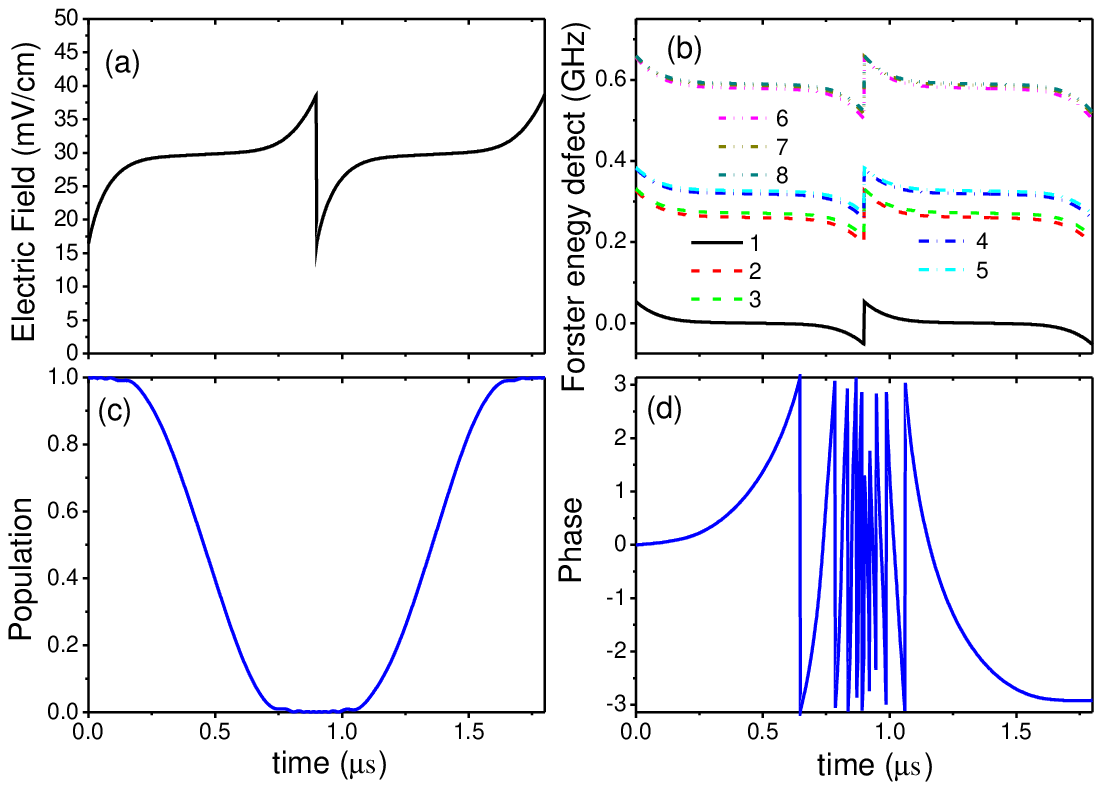}
\vspace{-.5cm}
\caption{
\label{Degenerate}(Color online) (a) Time dependence of the electric field for Stark tuning of the $\ket{90S_{1/2},96S_{1/2}} \to \ket{90P_{1/2} , 95P_{1/2}}$ F\"{o}rster resonance; (b) Time dependence of the energy defects of the interaction channels listed in Table~I for the $\ket{90S,96S} \to \ket{90P,95P}$ F\"{o}rster resonance; (c) Time dependence of the population of the collective state $\ket{90S_{1/2} ,96S_{1/2}}$; (d) Time dependence of phase of the collective state $\ket{90S_{1/2},96S_{1/2}}$.
}
\end{figure}
\end{center}

The time dependence of the electric field required to form the nonlinearly shaped detuning $\delta _{k} \left(t\right)=s_{1} \left(t-t_{k} \right)+s_{2} \left(t-t_{k} \right)^5$ of the $\ket{90S_{1/2} ,96S_{1/2}} \to \ket{90P_{1/2} ,95P_{1/2}}$ F\"{o}rster resonance with $s_1/2 \pi= -10\;\mathrm{MHz} / \mu s$ and $s_2 /2\pi =-2600\; \mathrm{MHz} /\mu s^{5}$, $t_1 =450$~ns and $t_2 =1350$~ns is shown in Fig.~\ref{Degenerate}(a). The time dependent F\"{o}rster energy defects for most important channels from Table~III are shown in Fig.~\ref{Degenerate}(b). The off-resonant excitation of various F\"{o}rster channels, partial overlapping of the resonances, and the finite lifetimes of Rydberg states are the most important limiting factors for quantum gate performance. 

\begin{table}
\caption{F\"{o}rster resonance channels $\ket{\gamma_a\,m_a,\,\gamma_b\,m_b}\to \ket{\gamma_{\alpha}\,m_{\alpha},\,\gamma_{\beta}\,m_{\beta}}$ and their angular factors. Here $m_a=m_b=1/2$.}
\begin{tabular*}{\columnwidth}{@{\extracolsep{\fill}}|c|c|c|c|c|c|c|c|} \hline 
 & $\ket{\gamma_a}$ & $\ket{\gamma_b}$ & $\ket{\gamma_{\alpha}}$ & $ \ket{\gamma_{\beta}}$ & $m_{\alpha}$ & $m_{\beta}$ & $Q_{k}$ \\ \hline 
1 & $90S_{1/2}$ & $96S_{1/2}$ & $90P_{1/2}$ & $95P_{1/2}$ & $1/2$ & $1/2$ & $-2/3$ \\ \hline 
2 & $90S_{1/2}$ & $96S_{1/2}$ & $90P_{1/2}$ & $95P_{3/2}$ & $1/2$ & $1/2$ & $-2\sqrt{2}/3$ \\ \hline 
3 & $90S_{1/2}$ & $96S_{1/2}$ & $90P_{1/2}$ & $95P_{3/2}$ & $-1/2$ & $3/2$ & $-\sqrt{2}/3$ \\ \hline 
4 & $90S_{1/2}$ & $96S_{1/2}$ & $90P_{3/2}$ & $95P_{1/2}$ & $1/2$ & $1/2$ & $-2\sqrt{2}/3$ \\ \hline 
5 & $90S_{1/2}$ & $96S_{1/2}$ & $90P_{3/2}$ & $95P_{1/2}$ & $3/2$ & $-1/2$ & $-\sqrt{2}/3$ \\ \hline 
6 & $90S_{1/2}$ & $96S_{1/2}$ & $90P_{3/2}$ & $95P_{3/2}$ & $1/2$ & $1/2$ & $-4/3$ \\ \hline 
7 & $90S_{1/2}$ & $96S_{1/2}$ & $90P_{3/2}$ & $95P_{3/2}$ & $-1/2$ & $3/2$ & $-1/\sqrt{3}$ \\ \hline 
8 & $90S_{1/2}$ & $96S_{1/2}$ & $90P_{3/2}$ & $95P_{3/2}$ & $3/2$ & $-1/2$ & $-1/\sqrt{3}$ \\ \hline 
\end{tabular*}
\end{table}

 The time dependence of the population [Fig.~\ref{Degenerate}(c)] and phase [Fig.~\ref{Degenerate}(d)] of the collective $\ket{ 90S_{1/2} ,96S_{1/2}}$ state for two interacting Rydberg atoms located at distance $R$=25~$\mu m$ along the \textit{z} axis was calculated taking into account all 116 interaction channels (examples are given in Table~II and Table~III). The off-resonant interaction channels lead to the undesirable phase shift, which is clearly seen in Fig.~\ref{Degenerate}(d). This shift can be partly compensated by adjusting the shape of the electric-field pulse, for example, by changing the time position of the second resonance to $t_{2} =1350.6$ ns. This value is sensitive to the accuracy of the calculated polarizability of Rydberg states.

\begin{center}
\begin{figure}[!t]
 \center
\includegraphics[width=\columnwidth]{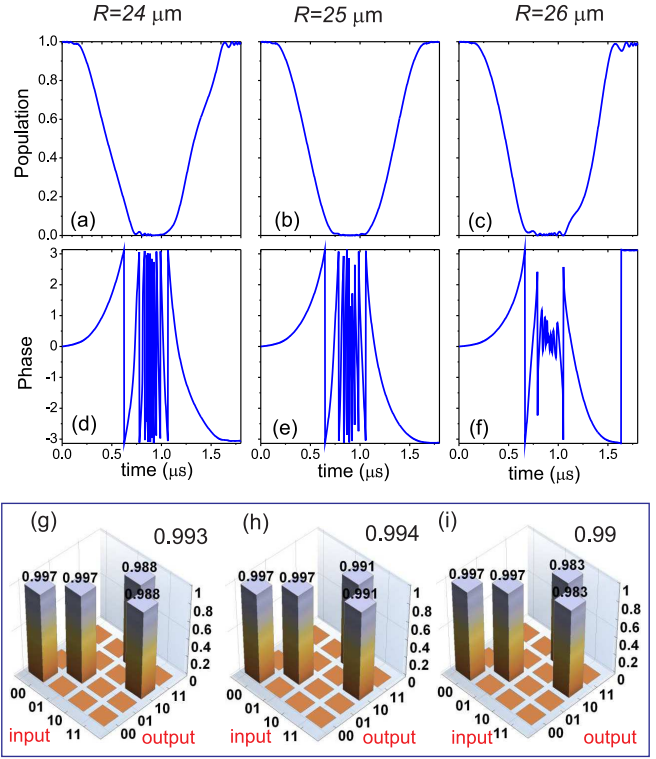}
\vspace{-.5cm}
\caption{
\label{Corr}(Color online) Double adiabatic passage of the Stark-tuned F\"{o}rster resonance for different interatomic distances \textit{R} with error correction.(a),(b),(c) Time dependences of population of the collective state $\ket{90S_{1/2},96S_{1/2}}$ calculated for \textit{R}=24, 25 and 26~$\mu \rm m$, respectively; (d),(e),(f) Time dependences of phase of the collective state $\ket{90S_{1/2} ,96S_{1/2}} $ calculated for \textit{d}=24, 25 and 26~$\mu \rm m$, respectively; (g),(h),(i) Calculated truth tables of a CNOT gate for \textit{R}=24, 25 and 26~$\mu \rm m$, respectively. The overlap with the ideal truth table is shown above each plot.
}
\end{figure}
\end{center}

With this correction we have calculated the time dependence of population and phase of the collective $\ket{90S_{1/2} ,96S_{1/2}} $ state for slightly different interatomic distances \textit{R}=24~$\mu \rm m$ (left-hand panel in Fig.~\ref{Corr}), \textit{R}=25~$\mu \rm m$ (central panel in Fig.~\ref{Corr}) and \textit{R}=26~$\mu \rm m$ [right-hand panel in Fig.~\ref{Corr}(b)]. Our calculations have shown that this variation of the interatomic distance leads to small phase changes at the end of the adiabatic passage, thus evidencing that our method to perform two-qubit quantum gates is insensitive to the atom position uncertainty. 

To estimate the fidelity of our schemes for two-qubit gates in realistic experimental conditions we have numerically calculated the truth table of a CNOT gate [Fig.~\ref{Scheme}] using a master equation for the density matrix and taking into account finite lifetimes of the Rydberg levels. We have also taken into account the phase shifts of the $\ket{90S_{1/2}}$ and $\ket{96S_{1/2}}$ Rydberg states in the time-dependent electric field by correcting the phase of the laser pulse~3 at Fig.~\ref{Scheme}(b) individually for control and target qubit.

We have solved the master equation
\begin{equation}
\label{Master}
\dot {\rho} \left( {t} \right) = - \frac{{i}}{{\hbar} }\left[ {\hat{H},\rho \left( 
{t} \right)} \right] + \hat {L}\left[\rho \left( {t} \right)\right]
\end{equation}
with \begin{equation}
\label{Decay}
\hat {L}[\rho] = \hat {L}^{(a)}[ \rho]+\hat {L}^{(b)} [\rho]. 
\end{equation}
The Liouvillian superoperator accounts for depopulation of the levels involved in the gate operation due to spontaneous and blackbody driven transitions to other Rydberg levels $\ket{r}$ and to low lying states. The superoperator also includes terms that repopulate the levels used for the gate, since blackbody driven transitions work in both directions between pairs of Rydberg states. For simplicity we have only included the terms that depopulate the Rydberg gate levels. This approximation slightly overestimates the gate errors thereby providing a conservative estimate of the gate fidelity. High $n$ Rydberg states decay to both neighboring Rydberg states and low lying states, with approximately 
equal rates in these two decay paths~\cite{Beterov2009}. Since both types of decay take the atom out of the computational subspace with high 
probability we have simply described the decay as being solely due to transitions to neighboring Rydberg states by setting
\be
\hat {L}^{(a,b)} \rho =- \frac{{1}}{{2}}\sum\limits_{r}{\gamma_{r} 
\left[ { \hat {\sigma} _{rr}^{(a,b)} \rho + \rho \hat {\sigma} _{rr}^{(a,b)}}  
\right]},
\ee
where   $\hat {\sigma} _{m n} ^{(a,b)} = {}_{(a,b)}\ket{m}\bra{n}_{(a,b)}$ is a transition operator for 
each of the atoms. The sum is taken over all Rydberg states $\ket{r}$ included in the simulations.

The truth tables were calculated for the same interatomic distances as previously (\textit{R}=24, 25 and 26~$\mu \rm m$). We have  found that for \textit{R}=25~$\mu \rm m$ the error is less than 1\%, and it only slightly increases when the distance between the atoms is varied. The main source of this error is revealed to be finite lifetimes of Rydberg atoms. In a 300~K environment the Rydberg states used have lifetimes~\cite{Beterov2009} 
$\tau_{90S}=270~\mu\rm s$, 
$\tau_{96S}=314~\mu\rm s$,
$\tau_{90P}=361~\mu\rm s$, 
$\tau_{95P}=406~\mu\rm s$.

\begin{center}
\begin{figure}[!t]
 \center
\includegraphics[width=\columnwidth]{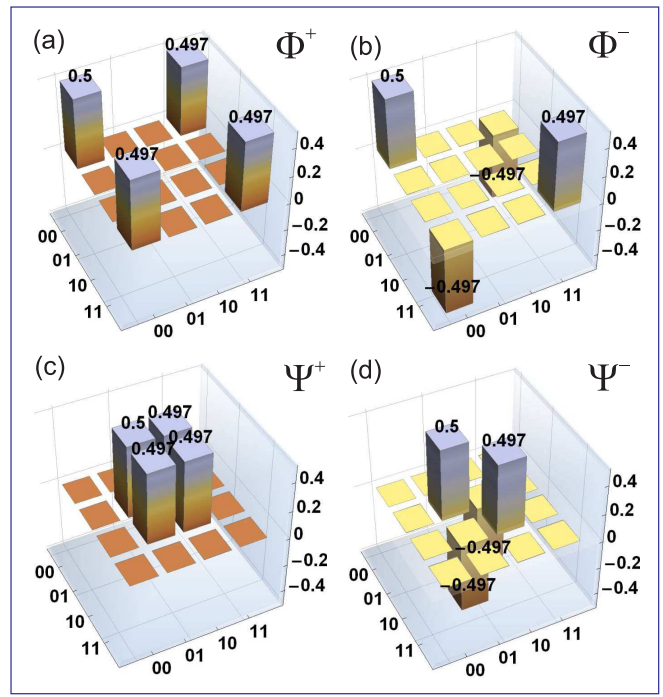}
\vspace{-.5cm}
\caption{
\label{Bell}(Color online) The reconstructed density matrices of the (a) $\Phi^+$, (b) $\Phi^-$, (c) $\Psi^+$ and (d) $\Psi^-$ Bell states.
}
\end{figure}
\end{center}

We have studied the phase errors by calculation of the fidelity of the Bell states which are  created by Hadamard gate applied to a control qubit, and a subsequent CNOT applied to a pair of qubits. The Bell states of a bipartite quantum system are defined as following:
\bea
\label{eq21}
\Phi^+&=&\frac 1{\sqrt 2}\left(\ket{00} +\ket{11} \right) \nonumber\\ 
\Phi ^-&=&\frac 1{\sqrt 2}\left(\ket{00} -\ket{11} \right) \nonumber\\
\Psi ^+&=&\frac 1{\sqrt 2}\left(\ket{01} +\ket{10} \right) \nonumber\\
\Psi ^-&=&\frac 1{\sqrt 2}\left(\ket{01}-\ket{10} \right).
\eea
\noindent  The density matrices of the generated Bell states after using maximum-likelihood reconstruction are shown in
Fig.~(\ref{Bell}) for interatomic distance $R=25\,\mu \rm m$. The calculated Bell state fidelities taking into account Rydberg lifetimes were better than 0.99. For $R=24\,\mu \rm m$ and $R=26\,\mu\rm m$ the fidelities are reduced to 0.965 and 0.984, respectively. Variation of the delay of the second resonance by 100 ps also reduces the Bell fidelities to 0.97 at $R=25\,\mu\rm m$.

Atomic qubits trapped in an optically defined array are subject to only small
position variations. For example using trap parameters from \cite{Maller2015a,Piotrowicz2013} 
and assuming an atom temperature of $T=10~\mu\rm K$ gives an in-plane position 
standard deviation of $\delta x \sim 0.1~\mu\rm m$ and an out of plane variation of 
$\delta z \sim 1.5~\mu\rm m$. This implies a variation around the $25~\mu\rm m$ 
nominal separation of $\pm0.15~\mu\rm m$. 
Comparing with Fig.~\ref{Corr} we anticipate less than 0.001 variation in gate 
fidelity for these conditions. Our calculations therefore show 
that the proposed gate protocol is insensitive to realistic 
experimental variations in the atom position.

\section{Summary}

We investigated the adiabatic passage across a F\"{o}rster resonance which can be considered as an alternative to the Rydberg blockade for implementation of  two-qubit quantum gates with Rydberg atoms. 

The gate fidelity has been found to be limited mainly by 
finite lifetimes of Rydberg states and dephasing due to 
off-resonant excitation of various F\"{o}rster interaction channels. We have shown, however, that only the limited number of channels affects the population dynamics of the initially excited collective state.
The decay of Rydberg population during the gate gives the radiative decay error 
of approximately $10^{-2}$ which is close to the calculated gate error in Fig.~\ref{Corr}(h) and Fig~\ref{Bell}. Reducing this error requires shorter gate times and larger separation from the neighboring F\"{o}rster resonances which can be observed only for the lower states with shorter lifetimes [see Fig.~\ref{Stark}(a)].  Although quantum gates based on Rydberg blockade in theory could provide the errors below $10^{-4}$~\cite{Theis2016}, such fidelity has not yet been demonstrated experimentally.

In contrast to the Rydberg-blockade gates, our approach does not require 
strong interaction between Rydberg atoms and can be potentially advantageous for implementing gates at large interatomic spacings. Although a $10^{-4}$ gate error is widely considered to be necessary for scalable quantum computation with a realistic
overhead in terms of the number of physical qubits the availability of long range gates with lower fidelity can be a 
useful feature of a large scale architecture. In order to move quantum information between qubits 
with large physical separation one can execute a string of swap gates  
using high fidelity local operations. The alternative is to use a lower fidelity gate that operates at long range to create 
Bell pairs with moderate fidelity, followed by entanglement purification with local operations~\cite{Bennett1996b}, 
and teleportation~\cite{Gottesman1999}. The gate protocol analyzed here provides CNOT truth table fidelity of $>0.99$ and creates maximally entangled Bell pairs with fidelity $>0.986$. With a qubit spacing of 
$\sim 4~\mu\rm m$ in a 2D array~\cite{Maller2015a} the $R=25~\mu\rm m$  range gate we analyze here would enable entanglement of 
arbitrary pairs in a block of 25 qubits suitable for encoding medium sized logical qubits. 

The F\"{o}rster resonances in time-varying electric field have been recently studied experimentally~\cite{Yakshina2016}. It has been shown that even for moderate interaction strengths it is possible to observe them on a short timescale of 100~ns.

This work was supported by the Russian Science Foundation Grant No. 16-12-00028 in the part of numeric simulation of the two-qubit gates and Bell states, by RFBR Grants No. 14-02-00680 and 16-02-00383, by Novosibirsk State University and Russian Academy of Sciences. MS was supported by NSF award 1521374, the AFOSR MURI on Quantum Memories and Light-Matter Interfaces, and the ARL-CDQI through cooperative agreement W911NF-15-2-0061. S.B.  was supported by EPSRC grant no EP/K022938/1.

%

\end{document}